# QDDS: A Novel Quantum Swarm Algorithm Inspired by a Double Dirac Delta Potential

Saptarshi Sengupta, Sanchita Basak, Richard Alan Peters II
Department of Electrical Engineering and Computer Science
Vanderbilt University, Nashville, TN, USA
E-mail: {saptarshi.sengupta, sanchita.basak, alan.peters}@vanderbilt.edu

*Abstract* – **In this paper a novel Quantum Double Delta Swarm (QDDS) algorithm modeled after the mechanism of convergence to the center of attractive potential field generated within a single well in a double Dirac delta well setup has been put forward and the preliminaries discussed. Theoretical foundations and experimental illustrations have been incorporated to provide a first basis for further development, specifically in refinement of solutions and applicability to problems in high dimensional spaces. Simulations are carried out over varying dimensionality on four benchmark functions, viz. Rosenbrock, Rastrigrin, Griewank and Sphere as well as the multidimensional Finite Impulse Response (FIR) Filter design problem with different population sizes. Test results illustrate the algorithm yields superior results to some related reports in the literature while reinforcing the need of substantial future work to deliver near-optimal results consistently, especially if dimensionality scales up.**

*Keywords— Quantum Particle Swarm; Swarm Intelligence; Double Delta Potential Well; Quantum Mechanics; Optimization*

## I. INTRODUCTION

The traditional Quantum-behaved Particle Swarm Optimization (QPSO) algorithms [1-3] (both Types I and II) extend the classical Newtonian dynamics of agent propagation in the canonical Particle Swarm Optimization (PSO) [4-5] [13] to a quantum framework. The convergence of particles to promising regions in the solution space in QPSO is driven by an attractive potential field directed towards the center of a singular Dirac Delta well. This is followed by the collapse of the wavefunction indicative of the particles' states using recursive Monte Carlo. The traditional types of QPSO are quite efficient and inexpensive candidates suitable for highly non-linear and non-convex optimization leveraging the quantum nature of the particles and the corresponding wavefunctions that sample a larger region of the solution space as compared to their classical counterpart employing a binary strategy: a particle is either present at a unique location or not. However, the intuitively simple rendering of the binding force exerted by a single delta well on a particular particle has been the subject of scrutiny and further research as demonstrated by [6]. One line of thought contends that the attractive coupling offered by a multi-well attractor is stronger than that offered by a singular one therefore facilitating a stronger stable equilibrium criterion [7]. Xie et al. [6] have recently proposed a quantum-behaved PSO based on a double delta model which assimilates the following three components: a) the global best (gbest) position, b) an agent's location with respect to the gbest position and c) an agent's location with respect to the mean of individual agents' best positions. The authors chose to model the personal and global best positions as centers of two singular delta potential wells, thereby arriving at a multi-scale representation of a double delta potential well. Simulations on widely used test functions such as Rosenbrock, Rastrigrin, Griewank and Sphere using varying population sizes for problem dimensionality 10 through 30 at a step size of 10 indicate the effective outcomes obtained using the algorithm. Further, the authors note that using two attractors in place of one increased the global search capability and convergence accuracy. In our work however, we are concerned more about finding the state equations of particles based on two spatially co-located delta potential wells. To this effect, an isolated system of double Dirac-delta potential wells and convergence to its centers are considered. The resulting iterative state updates mimic the trajectory of a bound particle (E<0) as it moves towards the lowest energy configuration viz. the point where the attractive potential is the least i.e. the center of the dominant well. By identifying constraints on the motion of the particles guided by their probabilistic nature of existence, an iterative scheme of convergence is put forward. The resulting algorithm (QDDS) is tested on a suite of benchmark functions which have many local minima, are bowl-shaped or valley-shaped as well as on the design optimization goal of a low-pass Finite Impulse Response (FIR) filter. Trial evaluations indicate the efficiency of the algorithm in finding solutions of acceptable quality with room for improvement both in terms of computational expense and finetuning of solutions. The organization of the paper is as follows: Section II discusses the physics governing the quantum mechanical model and sets up algorithmic foundations, Section III walks through the pseudocode with implementation details and Section IV reports test results. Section V briefly analyzes the outcomes followed by an analysis of the QDDS mechanism in Section VI with concluding remarks in Section VII.

## II. SWARM PROPAGATION USING A DOUBLE-DELTA POTENTIAL WELL

We start from the time-independent Schrodinger's wave equation which is stated as:

$$[-\frac{\hbar^2}{2m}\nabla^2 + V(r)]\psi(r) = E\,\psi(r) \qquad (1)$$

$\psi(r)$, $V(r)$, $m$, $E$ and $\hbar$ represent the wave function, the potential function, the reduced mass, the energy of the particle and Planck's constant respectively. Let us consider a particle in a double delta well, whose potential can be expressed as:

$$V(r) = -\alpha\{\delta(r+a) + \delta(r-a)\} \quad (2)$$

$\alpha$ expresses the strength of the well and $\{-a, a\}$ are the centers of the two wells. Considering the even solutions of the time-independent Schrodinger's equation (Eq. 1) and assuming $V=0$ at regions away from the centers of the two wells we get:

$$-\frac{\hbar^2}{2m}\frac{d^2}{dr^2}\psi(r) = E\psi(r) \quad (3)$$

The intention is to find solutions of the wave function $\psi$ in a double delta well setup for E<0 (bound states) in regions $\mathbb{R}1: r \in (-\infty, a)$, $\mathbb{R}2: r \in (-a, a)$ and $\mathbb{R}3: r \in (a, \infty)$. Assuming k to be equal to $(\sqrt{2mE}/\hbar)$, the even solutions can be expressed as [8]:

$$\psi_e(x) = \begin{cases} Ae^{-kr} & r > a \\ Be^{-kr} + Ce^{kr} & 0 < r < a \\ Be^{kr} + Ce^{-kr} & -a < r < 0 \\ Ae^{kr} & r < -a \end{cases} \quad (4)$$

To solve for the constants described in the above equation, we solve for the continuity of the wave function $\psi_e$ at $r = a$ and $r = -a$ and for the continuity of the derivative of the wave function at $r = 0$. Hence, we arrive at the equations for $\psi_e$ stated below [8]:

$$\psi_e(r) = \begin{cases} B(1 + e^{2ka})e^{-kr} & r > a \\ B(e^{-kr} + e^{kr}) & -a < r < a \\ B(1 + e^{2ka})e^{kr} & r < -a \end{cases} \quad (5)$$

We are not considering the solution of the odd wave function $\psi_o$ here as the existence of a solution is not guaranteed [8]. It is also interesting to note that the bound state energy in a double delta potential well is lower than that compared to a single delta potential well by approximately by a factor of $(1.11)^2 \approx 1.2321$ [7]. Next, if we consider the behavioral dynamics of a particle to be compliant with the Schrodinger wave equation i.e. Eq. (1), then we need to find the particle's probability density function for its behavioral characterization, which is given by the square of the magnitude of the wave function described in Eq. (5). Thus, we have to find $|\psi(r)^2|$. In order to say that there is a greater than 50% chance of finding a particle in the vicinity of the center of any of the potential wells, the following criterion must be satisfied [1]:

$$\int_{-|r|}^{|r|} \psi(r)^2 \, dr > 0.5 \quad (6)$$

where $-|r|$ and $|r|$ denote the two boundaries of the vicinity. Although we are trying to confine the particle in any one of the two potential wells, the wave function considered here is different from that considered in the case of a single potential well setup (as in traditional QPSO), This is due to the influence of the other well which is taken into consideration when deriving conditions for the confinement of a particle in a single well. Eq. (6) can also be written as:

$$\int_{-|r|}^{|r|} \psi(r)^2 \, dr = 0.5g \quad (1 < g < 2) \quad (7)$$

For ease of computation, we now consider that one of the wells is centered at 0. Solving for conditions of confinement of the particle in that well and computing $\int_{-|r|}^{|r|} (\psi(r)^2 \, dr$ for regions $\mathbb{R}2_{0-}: r' \in (-r, 0)$ by applying the second condition of Eq. (5) and $\mathbb{R}2_{0+}: r' \in (0, r)$ by applying the first condition of Eq. (5), we arrive at the equation below:

$$B^2 = \frac{kg}{e^{2kr} - 5e^{-2kr} + 4kr + 4} \quad (8)$$

We replace the denominator of the R.H.S. ($e^{2kr} - 5e^{-2kr} + 4kr + 4$) as $\delta$. Thus, we get:

$$\delta = e^{2kr} - 5e^{-2kr} + 4kr + 4 \quad (9)$$

Equating $B^2$ in L.H.S. of equation (8) for any two consecutive iterations (assuming it is a constant over iterations as it not a function of time) we get Eq. (10), Eq. (11) and Eq. (12):

$$\frac{g_{iter-1}}{e^{2kr_{iter-1}} - 5e^{-2kr_{iter-1}} + 4kr_{iter-1} + 4} = \frac{g_{iter}}{e^{2kr_{iter}} - 5e^{-2kr_{iter}} + 4kr_{iter} + 4} \quad (10)$$

$$\Rightarrow \frac{g_{iter-1}}{\delta_{iter-1}} = \frac{g_{iter}}{\delta_{iter}} \quad (11)$$

$$\Rightarrow \delta_{iter} = G \cdot \delta_{iter-1} \quad (0.5 < G < 2) \quad (12)$$

G is the ratio $(g_{iter}/g_{iter-1})$ and can vary from 0.5 to 2 since (1<g<2). To keep a particle moving towards the center of a potential well we find $\delta_{iter} \mid (0.5\,\delta_{iter-1} < \delta_{iter} < 2\,\delta_{iter-1})$. Thus, we find $\delta_{iter}$ at the current iteration based on $\delta_{iter-1}$ (found in the previous iteration) by adding or subtracting the gradient of $\delta_{iter-1}$ multiplied by a learning rate. The governing conditions of finding $\delta_{iter}$ depend on the relation of $\delta_{iter-1}$ with its version in the previous iteration, i.e., $\delta_{iter-2}$ as well as the sign of the gradient of $\delta_{iter-1}$ as described in Algorithm 1. The learning rate $\theta$ is designed as:

$$\theta = (1 - \epsilon)\left(\frac{max.\ iterations - current\ iteration}{max.\ iterations}\right) + \epsilon \quad (13)$$

$\epsilon$ is a small fraction between [0,1] set by the user. The learning rate $\theta$ decreases linearly from 1 to $\epsilon$ with the passage of iterations. Once we obtain $\delta_{iter}$, we back-solve Eq. (9) to retrieve $r_{iter}$ which denotes a particle's position as well as the potential solution for that particular iteration. We let $r_{iter}$, i.e. a particle's position in the current iteration maintain a component towards the best position found so far *(gbest)* along with its current solution obtained from $\delta_{iter}$. Subsequently, a cost function is computed with the solution $r_{iter}$ and if it is better than the best cost found thus far, the cost and the corresponding solution are stored in memory. This process is repeated over the total number of evaluations to find the overall best cost and best solution for the swarm. The complete procedure is described in Section III.

## III. QUANTUM DOUBLE DELTA SWARM ALGORITHM (QDDS)

In this section, we present the pseudocode of the Quantum Double Delta Swarm (QDDS) Algorithm.

---

**Algorithm 1.** Quantum Double Delta Swarm Algorithm

---

*Initialization Phase*

1: *Initialize k*
2: *Initialize a small constant λ randomly*
3: *Initialize maximum no. of iterations as maxiter*
4: *Initialize bestcost*
5: *for each particle*
6:     *for each dimension*
7:         *Initialize positions $r_1$ and $r_2$ for iterations 1 and 2*
8:     *end for*
9: *end for*
10: *Generate $\delta_1$ and $\delta_2$ from $r_1$ and $r_2$ according to Eq. (9)*
11: *Set iteration count iter=3*

*Optimization Phase*

12: **while** (iter<maxiter) and
    {($\delta_{iter-1} < 0.5 * \delta_{iter-2}$) or ($\delta_{iter-1} > 2 * \delta_{iter-2}$)}
13:   *Find learning rate θ using eq. (13)*
14:   *Select a particle randomly*
15:   **for** *each dimension*
16:     *if ($\delta_{iter-1} > 2 * \delta_{iter-2}$) and $\nabla\delta_{iter-1}>0$*
17:       $\delta_{iter} = \delta_{iter-1} - \theta * \nabla\delta_{iter-1} * \lambda$
18:     *elseif ($\delta_{iter-1} > 2 * \delta_{iter-2}$) and $\nabla\delta_{iter-1}<0$*
19:       $\delta_{iter} = \delta_{iter-1} + \theta * \nabla\delta_{iter-1} * \lambda$
20:     *elseif ($\delta_{iter-1} < 0.5 * \delta_{iter-2}$) and $\nabla\delta_{iter-1}<0$*
21:       $\delta_{iter} = \delta_{iter-1} - \theta * \nabla\delta_{iter-1} * \lambda$
22:     *elseif ($\delta_{iter-1} < 0.5 * \delta_{iter-2}$) and $\nabla\delta_{iter-1}>0$*
23:       $\delta_{iter} = \delta_{iter-1} + \theta * \nabla\delta_{iter-1} * \lambda$
24:     *end if*
25:   *end for*
26:   *Solve $r_{iter}$ from $\delta_{iter}$*
27:   *Generate a random number $\rho$ between 0 and 1*
28:   $r_{iter}=\rho * r_{iter} + (1-\rho) * r_{gbest}$
29:   *Compute cost using $r_{iter}$*
30:   **if** $cost_{iter}$<*bestcost*
31:     *bestcost = $cost_{iter}$*
32:     *best_solution = $r_{iter}$*
33:   *end if*
34:   *iter = iter + 1*
35: **end while**

---

## IV. EXPERIMENTS AND RESULTS

### A. Benchmark Functions

To evaluate the performance of the proposed algorithm, benchmark functions such as Rastrigrin, Rosenbrock, Sphere and Griewank have been considered. In addition to this, we test the algorithm on the multidimensional low pass FIR filter design problem using filter orders 10 and 20.

**Table 1. Benchmark Functions Considered for Testing**

| Function | Expression | Min |
|---|---|---|
| Rastrigrin | $f(x) = An + \sum_{i=1}^{n}[x_i^2 - A\cos(2\pi x_i)]$, A=10 | 0 |
| Rosenbrock | $f(x) = \sum_{i=1}^{n-1}[100(x_{i+1} - x_i^2)^2 + (1 - x_i)^2]$ | 0 |
| Sphere | $f(x) = \sum_{i=1}^{n} x_i^2$ | 0 |
| Griewank | $f(x) = 1 + \frac{1}{4000}\sum_{i=1}^{n} x_i^2 - \prod_{i=1}^{n}\cos(\frac{x_i}{\sqrt{i}})$ | 0 |

### B. Example Application: Design of High Dimensional Finite Impulse Response (FIR) Filters

This subsection outlines the design procedure of a multidimensional low-pass Finite Impulse Response (FIR) filter as proposed in [9-10]. The ideal filter response $H_d(e^{j\omega})$ and system transfer function $H(z)$ governing the design of the filter are given by Eqs. (14) and (15) respectively:

$$H_d(e^{j\omega}) = 1 \quad\quad 0 \leq \omega \leq \omega_p$$
$$= 0 \quad\quad \omega_s \leq \omega \leq \pi \quad\quad (14)$$

$$H(z) = \sum_{n=0}^{N} h(n)z^{-n} \quad\quad n = 0, 1, \ldots N \quad\quad (15)$$

Here $h(n)$ denotes the filter's impulse response and $N$ is the order of the filter having $N+1$ coefficients. The normalized passband and stopband edge frequencies are $\omega_p$ and $\omega_s$ respectively and $E_p$ and $E_s$ are errors in pass band and stop band given by Eqs. (16) and (17).

$$E_p = \frac{1}{\pi}\int_0^{\omega_p}(1 - H(\omega))^2 d\omega \quad\quad (16)$$

$$E_s = \frac{1}{\pi}\int_{\omega_s}^{\pi} H(\omega)^2 d\omega \quad\quad (17)$$

A cost function $\gamma$ of choice used in [9] is re-used for the minimization objective:

$$\gamma = \eta E_p + (1 - \eta)E_s \quad\quad 0 \leq \eta \leq 1 \quad\quad (18)$$

The main focus of the optimization routine is to find a balanced response that seeks to minimize the cost function $\gamma$, thereby minimizing the passband and stopband errors. The trade-off between reducing $E_p$ and $E_s$ is controlled by selection of a weight $\eta \in [0,1]$ as per the user's design condition.

## C. Simulation Results on the Benchmark Functions

### Table 2. Experimental Results for the Rosenbrock Function using 10 Independent Trials of QDDS

| P | Dim | PSO [3] | | QPSO [3] | | WQPSO [3] | | QDDS | | | |
|---|---|---|---|---|---|---|---|---|---|---|---|
| | | Iter | Mean ± St. Dev. | Iter | Mean ± St. Dev. | Iter | Mean ± St. Dev. | Iter | Mean ± St. Dev | Best | Worst |
| **20** | 10 | 1000 | 94.1276 ±194.3648 | 1000 | 51.9761 ±0.4737 | 1000 | 35.8436 ±0.2843 | 250 | **8.8912 ±0.0529** | 8.8224 | 8.9700 |
| | 20 | 1500 | 204.337 ±293.4544 | 1500 | 136.8782 ±0.6417 | 1500 | 62.7696 ±0.4860 | 375 | **19.0593 ±0.0991** | 18.9071 | 19.2355 |
| | 30 | 2000 | 313.734 ±547.2635 | 2000 | 157.4707 ±0.8287 | 2000 | 70.9525 ±0.4283 | 500 | **29.4457 ±0.0883** | 29.2827 | 29.5926 |
| **40** | 10 | 1000 | 71.0239 ±174.1108 | 1000 | 17.3177 ±0.1515 | 1000 | 16.9583 ±0.1336 | 250 | **8.9030 ±0.0875** | 8.7483 | 9.0642 |
| | 20 | 1500 | 179.291 ±377.4305 | 1500 | 54.0411 ±0.4210 | 1500 | 54.2439 ±0.3752 | 375 | **19.1421 ±0.0745** | 18.9883 | 19.2278 |
| | 30 | 2000 | 289.593 ±478.6273 | 2000 | 81.1382 **±0.0319** | 2000 | 57.0883 ±0.3437 | 500 | **29.3855 ±0.1321** | 29.1769 | 29.5359 |
| **80** | 10 | 1000 | 37.3747 ±57.4734 | 1000 | **7.5755 ±0.2708** | 1000 | 10.1650 ±0.2345 | 250 | 8.9213 ±0.0670 | 8.8569 | 9.0454 |
| | 20 | 1500 | 83.6931 ±137.2637 | 1500 | 32.9970 ±0.2068 | 1500 | 47.0275 ±0.3507 | 375 | **19.0663 ±0.1339** | 18.8227 | 19.2097 |
| | 30 | 2000 | 202.672 ±289.9728 | 2000 | 53.6422 ±0.2616 | 2000 | 51.8299 ±0.3103 | 500 | **29.4015 ±0.1079** | 29.2291 | 29.5685 |

### Table 3. Experimental Results for the Rastrigrin Function using 10 Independent Trials of QDDS

| P | Dim | PSO [3] | | QPSO [3] | | WQPSO [3] | | QDDS | | | |
|---|---|---|---|---|---|---|---|---|---|---|---|
| | | Iter | Mean ± St. Dev. | Iter | Mean ± St. Dev. | Iter | Mean ± St. Dev. | Iter | Mean ± St. Dev | Best | Worst |
| **20** | 10 | 1000 | 5.5382 ±3.0477 | 1000 | 4.8274 **±0.0015** | 1000 | 4.0567 ±0.0094 | 250 | **0.1214 ±0.0592** | 0.0425 | 0.2394 |
| | 20 | 1500 | 23.1544 ±10.4739 | 1500 | 16.0519 ±0.0414 | 1500 | 12.1102 **±0.0287** | 375 | **0.5776 ±0.0967** | 0.4765 | 0.7564 |
| | 30 | 2000 | 47.4168 ±17.1595 | 2000 | 33.7218 **±0.0114** | 2000 | 23.5593 ±0.0713 | 500 | **1.1709 ±0.2528** | 0.7499 | 1.5330 |
| **40** | 10 | 1000 | 3.5778 ±2.1384 | 1000 | 3.1794 **±6.033e-04** | 1000 | 2.8163 ±0.0083 | 250 | **0.1163 ±0.0577** | 0.0459 | 0.2628 |
| | 20 | 1500 | 16.4337 ±5.4811 | 1500 | 10.8824 **±0.0496** | 1500 | 9.7992 ±0.0628 | 375 | **0.5539 ±0.1300** | 0.3117 | 0.6847 |
| | 30 | 2000 | 37.2796 ±14.2838 | 2000 | 21.4530 ±0.0949 | 2000 | 17.4436 **±0.0034** | 500 | **1.1440 ±0.2768** | 0.6379 | 1.4135 |
| **80** | 10 | 1000 | 2.5646 ±1.5728 | 1000 | 2.2962 ±0.0130 | 1000 | 1.8857 **±0.0118** | 250 | **0.1397 ±0.0486** | 0.0543 | 0.2145 |
| | 20 | 1500 | 13.3826 ±8.5137 | 1500 | 7.8544 **±0.0011** | 1500 | 7.2855 ±0.0032 | 375 | **0.5312 ±0.1331** | 0.3059 | 0.7633 |
| | 30 | 2000 | 28.6293 ±10.3431 | 2000 | 15.9474 **±0.0198** | 2000 | 15.0255 ±0.0294 | 500 | **1.3041 ±0.2578** | 0.9302 | 1.7447 |

### Table 4. Experimental Results for the Sphere Function using 10 Independent Trials of QDDS

| P | Dim | PSO [3] | | QPSO [3] | | WQPSO [3] | | QDDS | | | |
|---|---|---|---|---|---|---|---|---|---|---|---|
| | | Iter | Mean ± St. Dev. | Iter | Mean ± St. Dev. | Iter | Mean ± St. Dev. | Iter | Mean ± St. Dev | Best | Worst |
| **20** | 10 | 1000 | 3.16e-20 ±6.23e-20 | 1000 | 1.3909e-41 ±1.4049e-43 | 1000 | **2.2922e-056 ±1.5365e-58** | 250 | 6.2437e-04 ±3.0752e-04 | 3.0163e-04 | 0.0014 |
| | 20 | 1500 | 5.29e-11 ±1.56e-10 | 1500 | 3.5103e-22 ±3.5452e-24 | 1500 | **2.9451e-40 ±2.8717e-42** | 375 | 0.0027 ±6.9011e-04 | 0.0019 | 0.0039 |
| | 30 | 2000 | 2.45e-06 ±7.72e-06 | 2000 | 5.3183e-14 ±5.3623e-16 | 2000 | **3.9664e-33 ±3.8435e-35** | 500 | 0.0067 ±0.0013 | 0.0046 | 0.0085 |
| **40** | 10 | 1000 | 3.12e-23 ±8.01e-23 | 1000 | 2.5875e-71 ±2.6137e-73 | 1000 | **5.5806e-80 ±5.6370e-82** | 250 | 7.3478e-04 ±2.3036e-04 | 4.9032e-04 | 0.0012 |
| | 20 | 1500 | 4.16e-14 ±9.73e-14 | 1500 | 3.7125e-42 ±3.7500e-44 | 1500 | **8.8186e-055 ±7.1785e-57** | 375 | 0.0028 ±8.3546e-04 | 0.0016 | 0.0041 |
| | 30 | 2000 | 2.26e-10 ±5.10e-10 | 2000 | 4.2369e-30 ±1.7009e-33 | 2000 | **5.4389e-44 ±2.4132e-45** | 500 | 0.0060 ±8.4108e-04 | 0.0048 | 0.0072 |
| **80** | 10 | 1000 | 6.15e-28 ±2.63e-27 | 1000 | 8.5047e-102 ±7.5974e-104 | 1000 | **4.7144e-106 ±4.7620e-108** | 250 | 5.1027e-04 ±1.4758e-04 | 3.1457e-04 | 7.5794e-04 |
| | 20 | 1500 | 2.68e-17 ±5.24e-17 | 1500 | 1.1542e-68 ±1.1585e-70 | 1500 | **2.5982e-74 ±2.6243e-76** | 375 | 0.0026 ±6.0137e-04 | 0.0020 | 0.0040 |
| | 30 | 2000 | 2.47e-12 ±7.16e-12 | 2000 | 2.2866e-49 ±2.3070e-51 | 2000 | **2.3070e-51 ±1.9125e-62** | 500 | 0.0055 ±8.8175e-04 | 0.0043 | 0.0069 |

**Table 5. Experimental Results for the Griewank Function using 10 Independent Trials of QDDS**

| P | Dim | PSO [3] | | QPSO [3] | | WQPSO [3] | | QDDS | | | |
|---|---|---|---|---|---|---|---|---|---|---|---|
| | | Iter | Mean ± St. Dev. | Iter | Mean ± St. Dev. | Iter | Mean ± St. Dev. | Iter | Mean ± St. Dev | Best | Worst |
| **20** | 10 | 1000 | 0.09217 ±0.0833 | 1000 | 5.5093e-04 ±0.0657 | 1000 | 5.6353e-04 ±5.5093e-04 | 250 | **8.0851e-05 ±3.3375e-05** | 2.3539e-05 | 1.2727e-04 |
| | 20 | 1500 | 0.03002 ±0.03255 | 1500 | **1.0402e-04 ±0.0211** | 1500 | 2.1318e-04 ±1.0402e-04 | 375 | 1.9821e-04 **±6.2856e-05** | 1.2262e-04 | 2.9547e-04 |
| | 30 | 2000 | 0.01811 ±0.02477 | 2000 | **1.2425e-04 ±0.0110** | 2000 | 2.1286e-04 ±1.2425e-04 | 500 | 2.5607e-04 **±6.4991e-05** | 1.6388e-04 | 3.5948e-04 |
| **40** | 10 | 1000 | 0.08496 ±0.0726 | 1000 | 1.6026e-04 ±0.0496 | 1000 | 0.0020 ±1.6026e-04 | 250 | **6.9932e-05 ±3.3276e-05** | 3.2751e-05 | 1.3565e-04 |
| | 20 | 1500 | 0.02719 ±0.02517 | 1500 | 1.7127e-04 ±0.0167 | 1500 | **1.6861e-04 ±1.7127e-04** | 375 | 1.9180e-04 **±4.7197e-05** | 1.1890e-04 | 2.7165e-04 |
| | 30 | 2000 | 0.01267 ±0.01479 | 2000 | 3.9088e-05 ±0.0085 | 2000 | **3.6762e-05 ±3.9088e-05** | 500 | 2.3775e-04 ±5.3165e-05 | 1.5328e-04 | 3.1541e-04 |
| **80** | 10 | 1000 | 0.07484 ±0.07107 | 1000 | 3.3744e-04 ±0.0327 | 1000 | 1.5281e-04 ±3.3744e-04 | 250 | **7.4205e-05 ±3.2774e-05** | 2.9087e-05 | 1.4314e-04 |
| | 20 | 1500 | 0.02854 ±0.0268 | 1500 | 4.1701e-04 ±0.0168 | 1500 | 3.2549e-04 ±4.1701e-04 | 375 | **1.8714e-04 ±5.6483e-05** | 1.1359e-04 | 2.5826e-04 |
| | 30 | 2000 | 0.01258 ±0.01396 | 2000 | **1.3793e-05 ±0.0106** | 2000 | 4.2231e-05 ±1.3793e-05 | 500 | 2.8736e-04 ±4.6883e-05 | 2.0340e-04 | 3.6768e-04 |

### D. Parameter Settings

We choose the constant $k$ to be 5 and $\lambda$ to be the product of a random number drawn from a normal distribution with $\mu = 0$ and $\sigma = 0.5$ and a factor of the order of $10^{-3}$. $\rho$ is a random number drawn between 0 to 1. The learning rate $\theta$ decreases linearly with iterations from 1 to 0.3. All experiments are carried out on two Intel(R) Core(TM) i7-5500U CPU @ 2.40GHz with 8GB RAM and one Intel(R) Core(TM) i7-2600U CPU @ 3.40GHz with 16GB RAM using MATLAB R2017a. 10 trials are carried out and results reported without any use of GPUs.

Tables 2 through 5 report performance of the QDDS algorithm on the Rosenbrock, Rastrigrin, Sphere and Griewank functions as well as compare and contrast with performances of PSO, QPSO and Weighted Mean Best QPSO (WQPSO) on the same benchmarks as reported by Xi et al. in [3]. All mean values and standard deviations listed in Tables 2,3,4 and 5 for the algorithms PSO, QPSO and WQPSO have been obtained from the work of Xi et al [3]. These have been used for a comparison of the performance of our algorithm (QDDS) on one-fourth the number of iterations with all other conditions for population and dimension remaining the same.

$P$ and $Dim$ represent the number of particles and the dimensionality of the functions. Figures 1 through 12 plot the convergence profiling of the above experiments on the stated benchmarks of orders 10, 20 and 30 over 10 independent trials. For purposes of brevity, results of simulations using only population size 20 are reported in this section. A detailed analysis can be found in Section V.

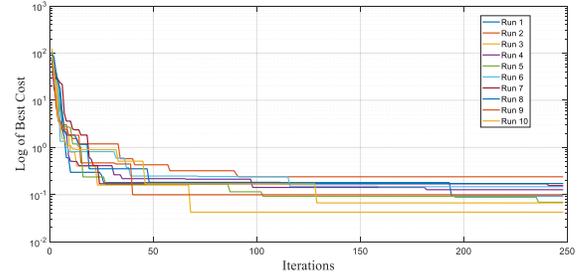

**Figure 1: Conv. Profiling of Rastrigrin using QDDS (dimension=10, population=20)**

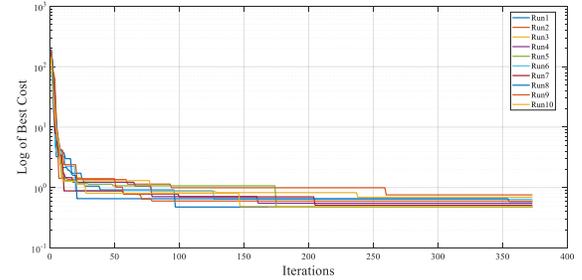

**Figure 2: Conv. Profiling of Rastrigrin using QDDS (dimension=20, population=20)**

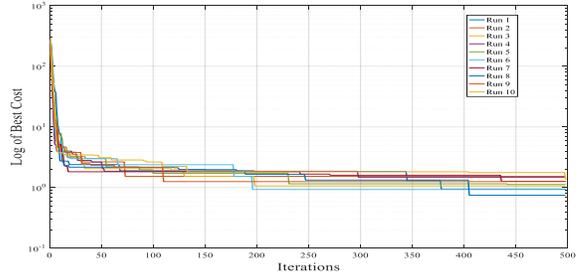

**Figure 3: Conv. Profiling of Rastrigrin using QDDS (dimension=30, population=20)**

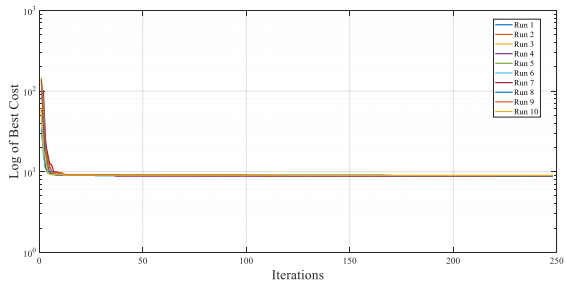

**Figure 4: Conv. Profiling of Rosenbrock using QDDS (dimension=10, population=20)**

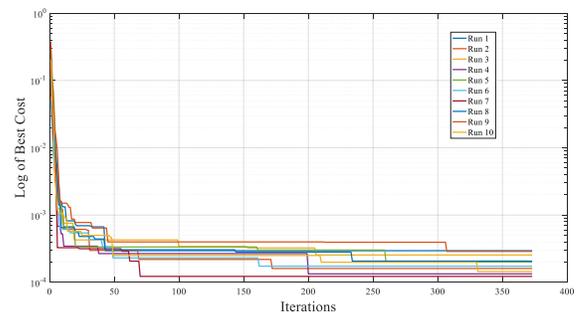

**Figure 8: Conv. Profiling of Griewank using QDDS (dimension=20, population=20)**

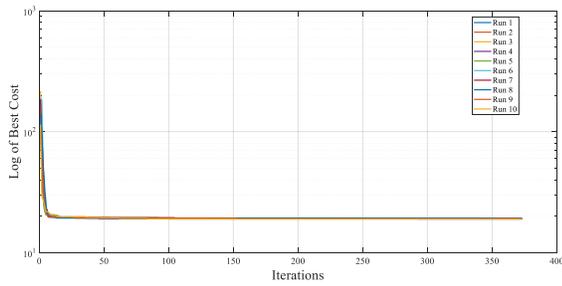

**Figure 5: Conv. Profiling of Rosenbrock using QDDS (dimension=20, population=20)**

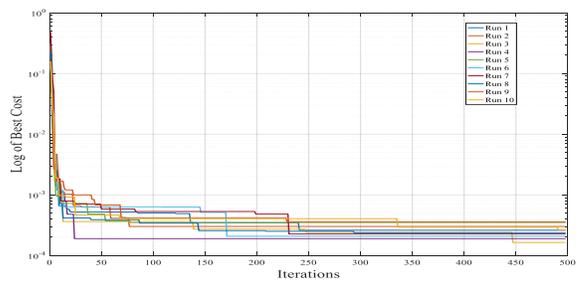

**Figure 9: Conv. Profiling of Griewank using QDDS (dimension=30, population=20)**

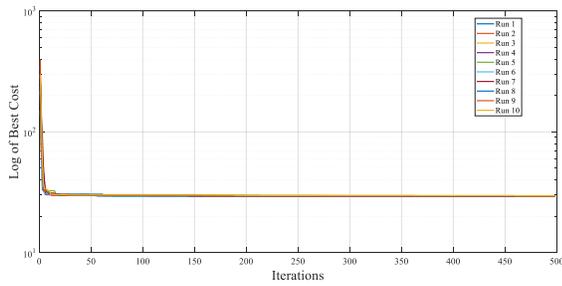

**Figure 6: Conv. Profiling of Rosenbrock using QDDS (dimension=30, population=20)**

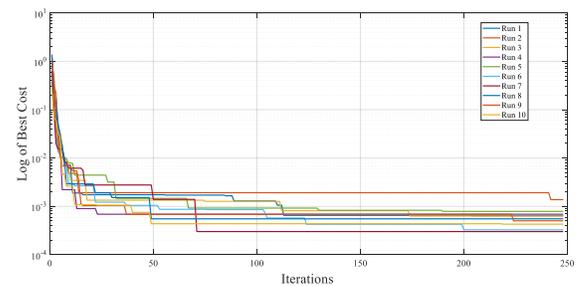

**Figure 10: Conv. Profiling of Sphere using QDDS (dimension=10, population=20)**

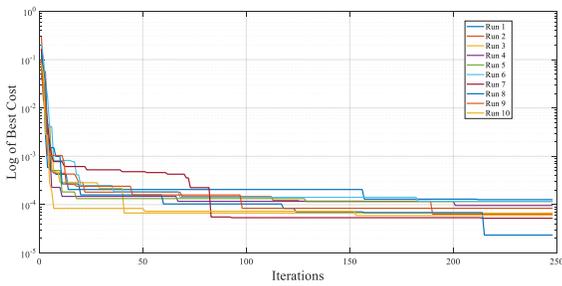

**Figure 7: Conv. Profiling of Griewank using QDDS (dimension=10, population=20)**

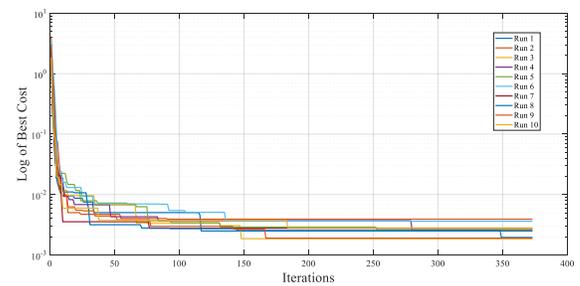

**Figure 11: Conv. Profiling of Sphere using QDDS (dimension=20, population=20)**

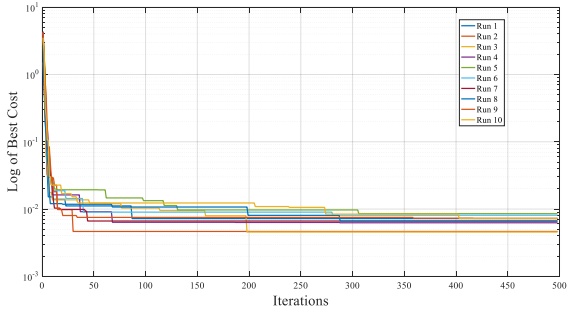

**Figure 12: Conv. Profiling of Sphere using QDDS (dimension=30, population=20)**

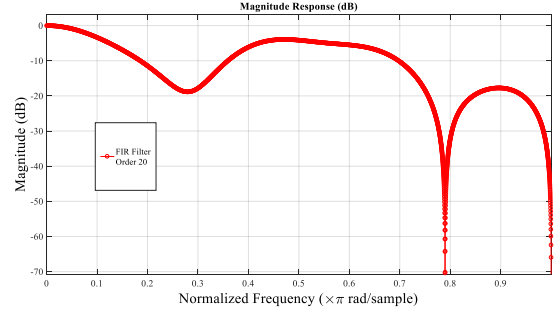

**Figure 14: Response of the 20-Dimensional FIR Filter**

*E. Simulation Results for the Finite Impulse Response (FIR) Filter Design Problem*

The following subsection illustrates the design of a multidimensional low-pass Finite Impulse Response (FIR) filter using a population size of 1000 and an iteration count of 250 and 500 for filter orders 10 and 20 respectively. The passband and stopband edges are set at $0.3\pi$ and $0.6\pi$.

**Table 6. Simulations for the 10-Dimensional FIR Filter Design Problem using 10 Independent Trials of QDDS**

| Δ (dB) | Mean Cost | St. Dev | Best Cost | Worst Cost |
|---|---|---|---|---|
| -13.6466 | 1.4817e-05 | 1.6762e-05 | 5.7632e-08 | 4.5367e-05 |

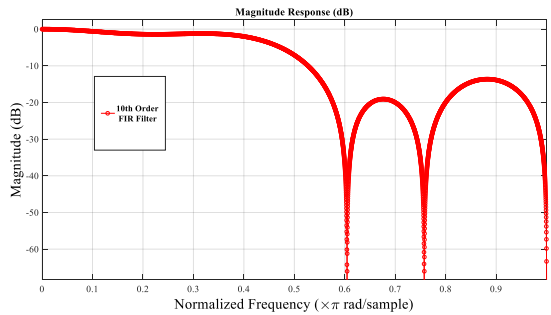

**Figure 13: Response of the 10-Dimensional FIR Filter**

**Table 7. Best 10-Dimensional Filter Coefficients (10 Trials)**

| Filter Coefficients | | |
|---|---|---|
| | $h(1) = h(10)$ | 0.070824792496751651 |
| | $h(2) = h(9)$ | -0.063184376757871669 |
| | $h(3) = h(8)$ | -0.038806613903081974 |
| | $h(4) = h(7)$ | 0.013227497402604124 |
| | $h(5) = h(6)$ | 0.39889122413816075 |

**Table 8. Simulations for the 20-Dimensional FIR Filter Design Problem using 10 Independent Trials of QDDS**

| Δ (dB) | Mean Cost | St. Dev | Best Cost | Worst Cost |
|---|---|---|---|---|
| -17.7398 | 7.1306e-05 | 9.9875e-05 | 1.6055e-06 | 3.1458e-04 |

**Table 9. Best 20-Dimensional Filter Coefficients (10 Trials)**

| Filter Coefficients | | |
|---|---|---|
| | $h(1) = h(20)$ | 0.011566963779404912 |
| | $h(2) = h(19)$ | 0.0077331878563942523 |
| | $h(3) = h(18)$ | -0.0094736298940968737 |
| | $h(4) = h(17)$ | -0.0068424142182682956 |
| | $h(5) = h(16)$ | 0.024047530227972496 |
| | $h(6) = h(15)$ | 0.040992486916104777 |
| | $h(7) = h(14)$ | 0.14983102243854188 |
| | $h(8) = h(13)$ | 0.0057626071427242216 |
| | $h(9) = h(12)$ | -0.0038505536917844913 |
| | $h(10) = h(11)$ | 0.28023279944300716 |

Tables 6 and 8 list the maximum stopband attenuation and the mean, standard deviation, best and worst cost values when using QDDS for 10 trials. Tables 7 and 9 report the set of best filter coefficients for orders 10 and 20 whereas Figures 13 and 14 plot the corresponding filter responses.

## V. ANALYSIS OF EXPERIMENTAL RESULTS

From Table 2 it is observed that the QDDS algorithm performs significantly better on the Rosenbrock function of dimensionality 10, 20 and 30 as compared to PSO, QPSO and WQPSO. In Table 3 the QDDS algorithm generates mean values which are at least 11.521 times smaller than WQPSO or even smaller in case of QPSO and PSO on the Rastrigrin function of dimensionality 10, 20 and 30. However, the standard deviation values obtained using WQPSO and QPSO are clearly superior to those found using QDDS. The results from Table 4 using the Sphere function indicate the sub-par performance of QDDS with respect to the competitor algorithms. Table 5 reports somewhat comparable results in terms of mean cost using WQPSO, QPSO and QDDS, while it is to be noted that QDDS has a standard deviation at least ~160 times smaller than that of QPSO. The convergence profiles in Figures 1 through 12 point out that QDDS is fairly consistent in its ability to converge to local optima of acceptable quality. It is obvious that the solutions to the problems discussed in the paper are in fact local optima, however the solution qualities corresponding to some of these local optima obtained using QDDS are evidently superior to some related reports in the literature [3,6,11,12]. One way to improve the performance of QDDS may be to not use gradient descent but a problem-independent optima seeking mechanism in the $\delta$ update step of the algorithm in Section III. The FIR filter responses record a maximum stopband attenuation of -13.6466 dB and -17.7398 dB on dimensions 10 and 20 for a QDDS implementation using gradient descent.

## VI. ANALYSIS OF THE QDDS MECHANISM

The Quantum Double Delta Swarm (QDDS) Algorithm is based on a quantum double Dirac delta potential well model and is an extension of the conventional QPSO (Type I and Type II) which are modeled after a singular Dirac delta potential well. The intuitively simple iterative updates of QDDS lead the swarm towards fitter regions of the search space in conjunction with reaching for regions of lower energy for a particle under the influence of a spatially co-located attractive double delta potential. The current form of the algorithm, however is prone to delivering suboptimal results because of the use of a gradient descent scheme in the $\delta$ update phase. In addition to this, the time complexity of the algorithm is markedly high due to the computationally heavy numerical approximation of $r_{iter}$ from $\delta_{iter}$ in the transcendental Eq. (9), as also outlined in Algorithm 1. The effects of using different social/cognition attractors as well as multi-scale particle topologies remain to be investigated and a thorough characterization of initialization schemes versus numerical accuracy is a logical follow-up. Overall, despite the high computational overhead QDDS produces acceptable solutions for some problems (Tables 2-3, 5) and not for some others, as seen in Table 4. However, the fact that QDDS is based on a legitimate optimization phenomenon in quantum physics and that it produces good quality solutions on some classical benchmarks warrants some resource expenditure in exploring better how it works.

## VII. CONCLUDING REMARKS

In this paper, a new swarming mechanism derived from the quantum mechanics of a double delta potential well is proposed and its inner workings explored. The mechanism (QDDS) is formally derived and a computable form is put forward followed by experiments to determine its accuracy in finding global optima. This is achieved by approaching some classic benchmark functions such as Rosenbrock, Rastrigrin, Sphere and Griewank as well as the multidimensional FIR filter design problem. Experimental results provide insight into the performance of the proposed approach on Rosenbrock, Rastrigrin and Griewank functions on which it appears to perform better, while not performing as well in the Sphere function. In addition to this, successful implementation of FIR filters of orders 10 and 20 are observed. While the present version of the algorithm is computationally expensive and stagnates occasionally, there is room for improvement in both areas. Future studies would aim at addressing these issues in addition to performing statistical significance tests to quantify the performance of the approach over an extensive suite of separable and non-separable functions of much higher dimensionality. However, QDDS captures a relatively unexplored approach in blending swarm intelligence and optimization and extends the literature on quantum-inspired computational intelligence.

## ACKNOWLEDGEMENTS

This work was made possible by the financial and computing support by the Vanderbilt University Department of EECS. The authors would also like to thank the anonymous reviewers for their suggestions on improving the paper.

## CONFLICT OF INTEREST

The authors declare that there is no conflict of interest regarding the publication of this paper.